\documentclass[12pt]{article}
\usepackage{times}
\usepackage{geometry}
\usepackage[utf8]{inputenc}
\usepackage{enumitem,amssymb}
\usepackage{ragged2e}
\usepackage{graphicx}
\usepackage{journalnames}
\usepackage{fancyhdr}
\usepackage{xcolor}
\usepackage[
    backend=biber, 
    natbib=true,
    style=numeric,
    sorting=none
]{biblatex}
\newlist{thematic}{itemize}{8}
\setlist[thematic]{label=$\square$}
\usepackage{pifont}

\begin{document}
\raggedright
\huge{
Standing on the Shoulders of Giants: A Comprehensive Spectroscopic Survey of Transiting \& High-Contrast Giant Planets}
\linebreak
\large

Munazza K. Alam, Emily Rickman, Kielan Hoch, Paul Molli\`ere, Josh Lothringer, Aarynn L. Carter, Isabel Rebollido, Ben J. Sutlieff, Jens Kammerer \\

\vspace{0.5cm}


\noindent \textbf{Thematic Areas (Check all that apply):} \linebreak \linebreak $\boxtimes$ (Theme A) Key science themes that should be prioritized for future JWST and HST observations 
\linebreak $\square$ (Theme B) Advice on optimal timing for substantive follow-up observations and mechanisms for enabling exoplanet science with HST and/or JWST \linebreak
$\square$ (Theme C) The appropriate scale of resources likely required to support exoplanet science with HST and/or JWST 
 \linebreak
$\boxtimes$ (Theme D) A specific concept for a large-scale ($\sim$500 hours) Director’s Discretionary exoplanet program to start implementation by JWST Cycle 3.
 \linebreak

\justify{
\textbf{Summary:} We propose a comprehensive survey of giant planets ranging from close-in highly irradiated hot Jupiters to young, wide-orbit directly imaged planets. The combination of two established techniques for probing planetary atmospheric compositions (time-series transit observations and high-contrast spectroscopy) will provide an unprecedented window into gaseous planet compositions across a range of equilibrium temperatures (100--2000 K), orbital separations (0.1--100 au), and system ages (10 Myr--1 Gyr). To-date, compositional measurements of both transiting and directly imaged planets suggest two distinct formation pathways: star-like formation for directly imaged planets and planet-like formation for transiting planets. By leveraging the combined technical and theoretical expertise of the transiting and direct imaging communities, we can obtain a holistic view of giant planet formation, migration, accretion, and evolution. From the results of this comprehensive program, we will begin to answer one of the fundamental outstanding questions in our understanding of giant planets: \textit{where and how in the protoplanetary disk do giant planets form?}


\pagebreak
\justify{

\textbf{Anticipated Science Objectives:}
 From the results of this comprehensive survey, we will: 1) measure the carbon-to-oxygen ratio, C/O, for temperate to highly-irradiated gas giants using high-resolution spectra between 3-5 $\mu$m taken with NIRSpec fixed-slit (for the transiting targets) and NIRSpec IFU (for the high-contrast targets); 2) assemble a spectral sequence of giant planets across temperatures to characterize the diversity of their chemical compositions (Figure~\ref{fig:spectral-sequence}); and 3) identify the dominant giant planet formation pathways from this legacy sample (Figure \ref{fig:CO_form}).


\textbf{Urgency}: Within the first year of JWST science operations, community efforts to analyze atmospheric observations of giant planets have yielded novel insights -- from the photochemical by-product SO$_2$ on the transiting exoplanet WASP-39b \cite{JTEC23,Alderson23} to the detection of silicate clouds in the directly imaged companion VHS 1256 b \cite{Miles23}. With JWST/NIRSpec, we can now obtain transit and high-contrast spectroscopy for giant planets at comparable wavelengths and spectral resolutions \cite{Alderson23,Miles23}. Exploring these synergies early on in JWST's lifetime has the potential to impact future observational strategies with this facility.

\textbf{Risk/Feasibility}: Comparing transiting and directly imaged planets has not yet been thoroughly tested, so there is inherent risk is starting a large-scale observational campaign with an undemonstrated strategy. However, if this program is successful, it has the potential to elucidate giant planet formation and evolution -- a critical blindspot in our understanding of the planets most amenable to atmospheric characterization. 



\textbf{Timeliness}: Our strategy for combining the expertise of the transiting and high-contrast communities to unlock the formation pathways of giant planets dovetails nicely with the Astro2020 recommendation to prioritize exoplanet atmospheres and the origins of planetary systems \cite{Astro2020}.  
By testing this approach to combine high-contrast and transit observations early on in JWST's lifetime, we will explore avenues to pave the legacy of JWST and future science missions.

\textbf{Cannot be accomplished in the normal GO cycle}: While our strategy to combine transit and high-contrast spectroscopy observations is a scientifically valuable pathfinder proposal, the target selection requires more observationally expensive targets (e.g., long-period transiting exoplanets and high-contrast targets that are closer to their host star than typical direct spectroscopy targets) -- which might be harder to justify to a TAC. The program also requires a significant time investment with a sample large enough for a population study, as well as limited opportunities for repeat observations.

\newpage


\begin{figure}
    \centering
    \includegraphics[width=0.75\textwidth]{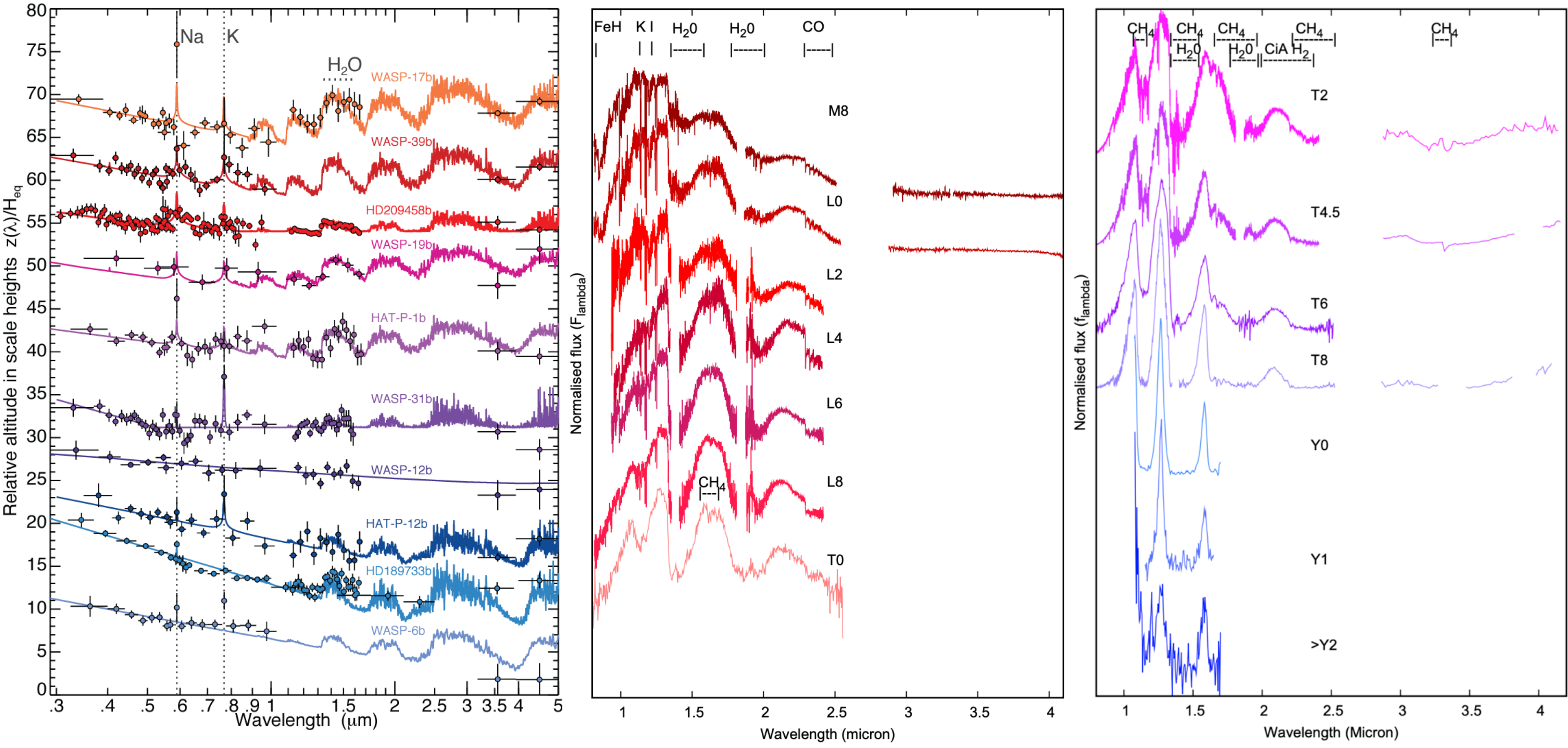}
    \caption{Spectral sequence of transiting exoplanets \cite{Sing16} and directly imaged brown dwarfs \cite{Helling14}.}
    \label{fig:spectral-sequence}
\end{figure}

\begin{figure}
    \centering
    \includegraphics[width=0.75\textwidth]{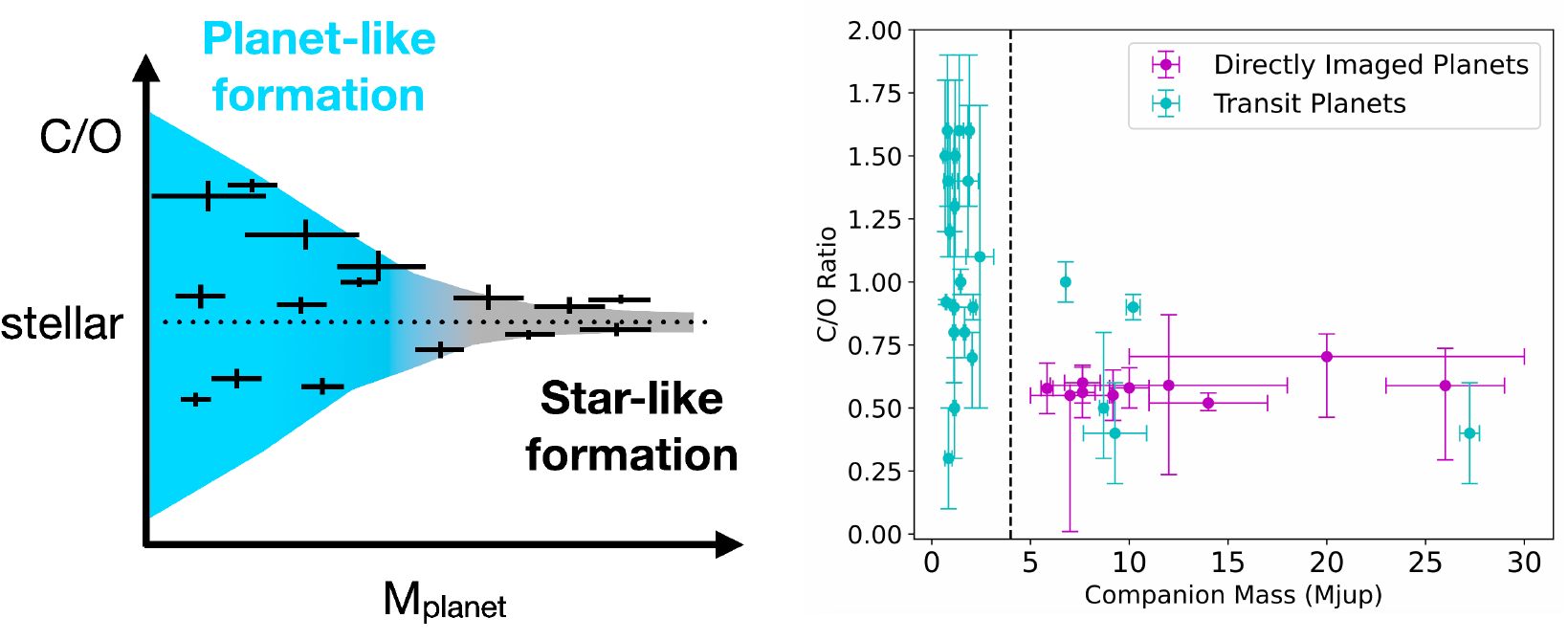}
    \caption{C/O measurements for low-mass transiting exoplanets and high-mass directly imaged planets suggest two distinct formation pathways for these populations of giant planets.}
    \label{fig:CO_form}
\end{figure}


\printbibliography
\end{document}